\documentstyle[editedvolume,phase1]{crckapb} 
\input{psfig}

\begin{opening}

\title{The Hubble Space Telescope Key Project \protect\\
         to Measure the Hubble Constant$^1$}

\author{Wendy L. Freedman}
\institute{Carnegie Observatories\\
           813 Santa Barbara St., Pasadena CA 91106}

\author{Jeremy R. Mould}
\institute{Mount Stromlo and Siding Spring Observatories\\
           Australian National University, Weston Creek, Australia}

\author{Robert C. Kennicutt Jr.}
\institute{Steward Observatory\\
           University of Arizona, Tucson AZ 85721}

\author{Barry F. Madore}
\institute{IPAC/Caltech\\
           100-25, Pasadena, CA 91125}

\end{opening}

\runningtitle{H0 Key Project}

\begin{document}


\section{Introduction}

A Joint Discussion on the  extragalactic distance scale and the Hubble
constant took   place fifteen years ago, at   the 1982 XVIIIth General
Assembly of the IAU, held in Patras, Greece.  At that time, the newest
applications   of infrared photometers   to Tully-Fisher  measurements
(Aaronson 1983)  and  Cepheid distances (Madore  1983)  were reported.
CCDs were just  coming   into use and had    not yet been  applied  to
extragalactic  distance determinations;    all of  the   extragalactic
Cepheid     distances   were    based  on    photographic   Argelander
(eye-estimated) photometry  (Tammann  and Sandage 1983 and  references
therein).   No  Cepheid distances to  type  Ia supernova-host galaxies
were available.

\bigskip

$^1$ {\it Based  on  an  Invited  Review  given at the  IAU  Symposium 183,
Cosmological Parameters and   the Evolution of  the Universe,  held in
Kyoto, Japan, August 1997.} 

\medskip

What is the  situation in 1997  at the time  of  IAU Symposium  183 in
Kyoto?  Since Patras, we have seen  a steady increase in the precision
with  which  extragalactic   distances  can be   measured.   With  the
widespread availability  of  linear   array  detectors,  more accurate
distances   to  both the  primary  and   secondary distances are being
obtained  ({\it  e.g.,} Jacoby {\it et   al.} 1992; Freedman  \& Madore
1996; Donahue   \& Livio  1997).    In parallel  to the   advances in
detector  technology has   been  the development  of  several  new and
independent techniques for measuring distances.  For the first time in
the history of this  difficult field,  relative distances to  galaxies
are  being  compared on a  case-by-case  basis, and their quantitative
agreement is  being established.  Briefly,  we review here progress on
the Hubble Space Telescope Key Project to measure H$_0$.

\section{The H$_0$ Key Project}

Converging on an accurate value of the Hubble constant has been a slow
and incremental process.  The  difficulties have stemmed  largely from
the effects of systematic errors in  the extragalactic distance scale.
For this reason, the Key Project has been designed to incorporate many
independent cross-checks of  both  the primary and  secondary distance
scales.  Rather than concentrate on one particular method, the goal of
the Key Project is  to  undertake a  comparison and a  calibration  of
several different  methods so that cross-checks  on  both the absolute
zero point as well as relative distances,  and therefore on H$_0$, can
be obtained.

Ultimately, the aim  of the Key Project is  to derive a value for  the
expansion rate of the Universe, the Hubble constant, to an accuracy of
10$\%$  (Freedman {\it  et al.} 1994a;  Kennicutt,   Freedman \& Mould
1995; Mould {\it   et al.} 1995; Madore  {\it  et al.} 1998).   It has
been designed with three primary goals: (1)  to discover Cepheids, and
thereby measure  accurate distances to  spiral galaxies located in the
field  and in small  groups that are  suitable  for the calibration of
several  independent  secondary methods,   (2)  to make direct Cepheid
measurements of distances  to three  spiral galaxies  in each of   the
Virgo and Fornax clusters,  and (3) to  provide  a check on  potential
systematic errors both in the Cepheid distance scale and the secondary
methods.  We briefly review the progress in each of these areas.

\section{ Measurement of Cepheid  Distances / Calibration of Secondary Methods}

To date  the  H$_0$ Key Project results  have  been published  for M81
(Freedman  {\it et al.}  1994b), M100  (Ferrarese {\it  et  al.} 1996;
Freedman {\it et  al.} 1994a), M101 (Kelson {\it  et al.} 1996, 1997),
NGC~925 (Silbermann  {\it et al.} 1996),  and NGC~3351 (Graham {\it et
al.} 1997).  Recently, we  have also determined distances  to NGC~3621
(Rawson {\it  et al.}  1998),   NGC~2090 (Phelps {\it  et al.}  1998),
NGC~7331 (Hughes {\it et al.}   1998), NGC~4414 (Turner {\it et   al.}
1998), and   NGC~1365 (Silbermann {\it et al.}   1998;  Madore {\it et
al.}   1998).  Significant  progress  has also  been  made in  the HST
supernova calibration project; Cepheids  have been located and studied
in IC~4182 (Saha  {\it  et al.} 1994), NGC~5253   (Saha  {\it et  al.}
1995) and NGC~4536  (Saha {\it et  al.} 1996).  Results have also been
published for NGC~4639 and NGC~4496A (Sandage {\it  et al.} 1996), and
Cepheids have   been detected in the   Leo I galaxy  NGC~3368 (M96) by
Tanvir {\it et al.} (1995).

%

\smallskip

To minimize the risk of systematic errors in the data reduction phase,
all of the reductions within the Key Project  effort are undertaken by
two independent groups, using two  different software packages: DoPHOT
and ALLFRAME (Schechter  {\it et al.}  1993;  Saha {\it et  al.} 1994;
Stetson  1994).  In addition, we are  currently performing a series of
artificial star tests to better quantify the  effects of crowding, and
to understand the limits in each of  these packages (Ferrarese {\it et
al.}, 1998 in  preparation).   Because  our requirement  for  an accurate,
absolute calibration    is critical,   we  are also     undertaking an
extensive, independent calibration  of  the WFPC2 zero point  (Stetson
{\it et  al.}, in  preparation), complementary to   the efforts of the
WFPC2 instrument team and the Space Telescope Science Institute.


Determination  of   H$_0$  to an  accuracy   of  10$\%$  requires that
measurements be acquired at great enough distances and in a variety of
directions so that the average  contribution from peculiar motions  of
galaxies  is significantly less than 10$\%$   of the overall expansion
velocity.  The current limit  for detection of Cepheids  with HST is a
distance of about 25-30  Mpc (Newman {\it et  al.} 1998; Saha  {\it et
al.} 1996), where peculiar motions can still contribute 10-20\% of the
observed velocity.  Hence, the  main thrust of  the Key Project is the
calibration of secondary distance indicators which then operate out to
distances  significantly greater than  can  be measured with  Cepheids
alone.  The calibrating galaxies in the  sample all have velocities of
less   than $\sim$1,500 km/sec.  Even  with  the relative proximity of
these   galaxies,   discovering   Cepheids    remains a   challenging,
time-consuming  task  using HST; the  integration   times for the more
distant galaxies in the sample  can each amount  to over 30 orbits  of
HST time.

With the database of Cepheid distances  being assembled as part of the
H$_0$ Key  Project, a number of  secondary  indicators can be directly
calibrated and  tested.  Several of  these  methods can be applied  to
velocity-distances of 10,000 km/sec or greater.  These include type Ia
supernovae,  type II  supernovae, the  Tully-Fisher  relation, and the
D$_n$-$\sigma$ relation.   Applicable at intermediate distances is the
surface-brightness fluctuation method  ({\it e.g.,} Tonry {\it et al.}
1997).  Although the planetary nebula luminosity function method ({\it
e.g.,} Feldmeirer, Ciardullo \& Jacoby 1997) only extends over the same
range as  the  Cepheids (out  to about  20 Mpc), it  offers a valuable
comparison  and  test of  methods that operate  locally  (Cepheids, RR
Lyrae stars, tip   of  the red giant   branch (TRGB))  and those  that
operate  at   intermediate   and   greater  distances   ({\it    e.g.,}
surface-brightness fluctuations  and the  Tully-Fisher  relation). The
database of Cepheid distances will also provide a means for evaluating
as yet  less well-tested methods;  for instance, the  globular cluster
luminosity function  (for a recent application  see Baum {\it  et al.}
1997), red supergiants, and HII region diameters.

In  the limited space  available here, we confine   our remarks to the
calibration of the Tully-Fisher relation and type Ia supernovae.

\subsection{ Calibration of the Tully-Fisher Relation }

One of the key elements  of the HST  H$_0$ Key Project is the  Cepheid
calibration  of the   relation between the   luminosity and rotational
velocity of  spiral galaxies, the Tully-Fisher  (TF) relation.  On the
basis of their relatively   high inclinations, line widths, and   late
morphological types, fourteen of our target galaxies were chosen to be
useful as Tully-Fisher  calibrators.    These include NGC 3031,   925,
3351, 2090, 7331, 3621, 2541, 3198, 3319,  4725, 4535, 4548, 1365, and
1425.  Along with  NGC 3368 (Tanvir {\it et  al.} 1995), NGC 4536  and
4639  observed by Sandage and collaborators,  and the TF galaxies that
have  had Cepheid distances determined  from the ground, NGC 598, 224,
2403 and 300 (Freedman 1990;  Madore \& Freedman  1991), this yields a
total of 21 individual TF calibrators.  The new HST distances increase
by a  factor of 4 the  numbers of TF calibrators  previously available
from ground-based Cepheid searches.

The status of the H$_0$ Key  Project Tully-Fisher calibration has been
reviewed  recently by Mould {\it   et  al.} (1997).  This  preliminary
calibration yields  a value of H$_0$  =  73 $\pm$10  km/sec/Mpc.  This
value is in very good agreement with  a recent analysis of 24 clusters
with  Tully-Fisher measurements  by  Giovanelli {\it  et  al.} (1997).
Based on a similar set of Cepheid distances,  these authors find H$_0$
= 69 $\pm$ 5 km/sec/Mpc.

\begin{figure}
\psfig{figure=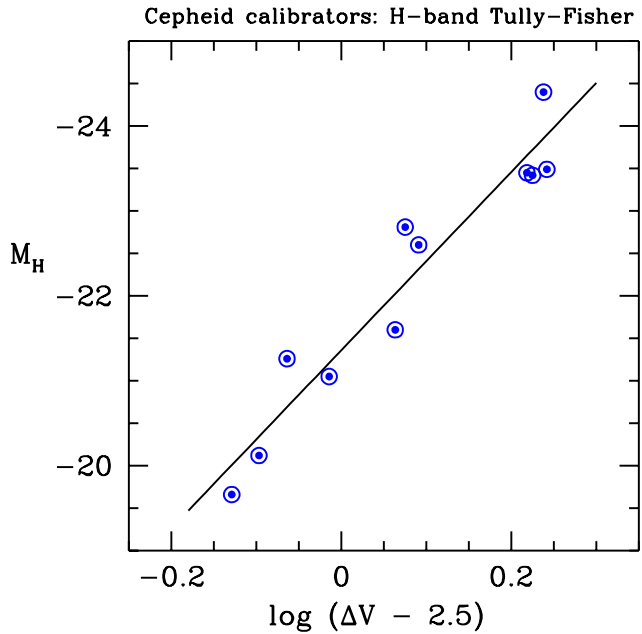,height=3.5in,angle=0}
\caption{ H-band Tully-Fisher relation for galaxies with distances
determined from Cepheid variables.}
\end{figure}


For illustration, we show in  Figure 1 an H-band Tully-Fisher relation
for the calibrating galaxies with measured Cepheid distances available
to date.  The H-band photometry and line widths are from Aaronson {\it
et al.} (1982);  these data (including  lower  line-width systems) are
presented by Mould {\it et al.} (1997).  Distances to the galaxies are
from Freedman (1990),  Freedman {\it et al.}   (1994), Tanvir {\it  et
al.} (1995), Silbermann   {\it et al.}   (1996), Graham  {\it  et al.}
(1997), Rawson {\it et  al.} (1997), and  Hughes {\it et al.}  (1998).
There are 11 galaxies plotted.
As  part of the Key Project, we
have obtained and are currently analyzing new  UBVRIJHK images of all
of the calibrating galaxies for calibration purposes.


\subsection{ Calibration of Type Ia Supernovae}

One of the most promising methods for  measuring relative distances to
distant galaxies is  based on the  measurement  of type  Ia supernovae
luminosities. Cepheid calibrators  have recently become available  for
this  method  as a  result of  the  availability  of  HST ({\it e.g.,}
Sandage {\it et al.} 1995 and references therein). Several independent
studies  now suggest   that  type Ia  supernovae  are  not  the simple
standard  candles that they  were earlier  suggested to  be,  but they
appear to obey  a  fairly well-defined  relation between the  absolute
magnitude at  maximum light and   the shape or    decline rate of  the
supernova light  curve (Phillips 1993; Hamuy  {\it et al.} 1995, 1996;
Reiss, Press \& Kirshner 1995).

As part of the  Key Project target  sample, we have observed  NGC 4414
(Turner {\it et al.} 1998,  in preparation), an inclined spiral galaxy
useful   as both  a Tully-Fisher  calibrator   as  well as  a  type Ia
supernova calibrator. It was   host to supernova 1974G;  unfortunately,
however, the   light curve  of   this supernova is   of only  moderate
quality.  The Fornax cluster elliptical galaxies NGC~1316 and NGC~1380
are each hosts  to the   well-observed  type Ia supernovae 1980N   and
1992A,  respectively.   (The supernova  1981D   was  also observed  in
NGC~1316, but the data are photographic, and hence are  not of as high
quality as the other two.)

The new Cepheid  distance to NGC~1365,  and associated estimate of the
distance   to  the Fornax cluster (discussed   below)   thus allow two
additional very high-quality objects to be added to the calibration of
type Ia supernovae. Including the the Fornax cluster supernovae (1980N
and 1992A),  in  addition to   the other  Cepheid calibrators of   the
Sandage {\it  et al.} program (1937C,  1972E, 1960F, 1981B, 1989B, and
1990N),    and  applying this  calibration   to   the  distant type Ia
supernovae of Hamuy  (1995) gives H$_0$ =  64--68~km/sec/Mpc (Freedman
{\it et al.}  1997).  The  larger value  of H$_0$ compared  to that of
Sandage {\it et al.} (1996)  (57 km/sec/Mpc) is  due to three factors:
(1)  we have  given   low  weight to  historical  supernovae  observed
photographically,     (2)   we     have   included    a   decline-rate
absolute-magnitude   relation,  and (3)  we have  added   2 new Fornax
calibrators.  All   three  factors contribute  in  roughly  comparable
proportions.   We note that  the   addition of the Fornax  calibrators
changes the value of H$_0$ by +3 km/sec/Mpc  or less than 5\%; most of
the  remaining difference reflects   the   lower weight given to   the
historical supernovae 1895B, 1937C, and 1961F.

\section{ The Virgo/Fornax clusters and the Local Cepheid Calibration }

The current  limit for the detection  of Cepheids with  HST appears to
have been reached at a velocity of  $\sim$3,000 km/sec (Newman {\it et
al.} 1998).  Hence, the existence of large-scale flows still precludes
the  current measurement  of  H$_0$ to  $\pm$10\% with Cepheids alone.
However, direct Cepheid distances to  the Virgo and Fornax clusters at
cz$\geq$1,200 km/sec can still provide  a consistency check at a level
of $\pm$20\%.


The Virgo cluster is not an ideal cluster for either the determination
of  H$_0$ or  the calibration  of secondary methods.   As discussed in
detail  in Freedman  {\it   et  al.} (1994a),   one  of  the  dominant
uncertainties in the determination of H$_0$ based on the Virgo cluster
is due to  the fact that  the distribution of  its spiral galaxies  is
both extended   and complex.  As such, a   single galaxy  alone cannot
define the mean distance to the Virgo cluster to an accuracy of better
than  15--20$\%$ (Freedman {\it  et  al.} 1994a,  Mould  {\it  et al.}
1995).  Cepheid distances to five spiral  galaxies in the Virgo galaxy
have now been published, and they are listed in  Table 1.  Despite all
of the complications, it is interesting that the mean Cepheid distance
agrees   very well with  recent  independent   estimates of the  Virgo
cluster distance obtained by Jacoby {\it et al.} (1997) and Tonry {\it
et al.} (1997) for elliptical galaxies.

Adopting a recession  velocity for the Virgo  cluster of 1,404 $\pm$80
km/sec (Huchra 1988) and a Virgo  distance of 17.8  Mpc yields a value
of  H$_0$ =   79  $\pm$6 (random)    $\pm$16 (systematic)  km/sec/Mpc.
Alternatively,  adopting a recession  velocity of 1,179 $\pm$17 km/sec
(Jerjen and Tammann 1993) results in H$_0$ = 66 $\pm$14 km/sec/Mpc for
the same  distance.   The  dominant sources   of uncertainty in   this
estimate are systematic: (a) reddening corrections, (b) the zero point
of the  Cepheid PL relation, (c)  the position  of these galaxies with
respect to  the  center  of the Virgo   cluster,  and (d) the  adopted
recession velocity of the cluster.

In addition to  the galaxies listed in  Table 1, two  additional Virgo
cluster galaxies (NGC 4535 and NGC 4548), have  been observed with HST
and are currently being analyzed as part of the Key Project.  However,
despite the increase in numbers of Virgo cluster galaxies with Cepheid
distances, the clumpiness,  large angular extent, and unknown peculiar
motion of the Virgo cluster all preclude a determination of the Hubble
constant at  a level better than $\sim  \pm$20$\%$.  A  more favorable
cluster for allowing a  consistency check of H$_0$ determinations from
secondary distance indicators is the Fornax cluster.

\medskip

\begin{table}[htb]
\begin{center}
  \caption{ CEPHEID DISTANCES TO VIRGO CLUSTER GALAXIES }
\begin{tabular}{lcc}
\hline
  Galaxy & Distance Modulus & Distance  \\[3pt]
\hline
 NGC 4321  & 31.04 $\pm$ 0.21 & 16.1 $\pm$ 1.5  \\
 NGC 4496A  & 31.13 $\pm$ 0.10 & 16.8 $\pm$ 0.8  \\
 NGC 4571  & 30.87  $\pm$ 0.15 & 14.9 $\pm$ 1.2  \\
 NGC 4536$^1$  & 31.10 $\pm$ 0.13 & 16.6 $\pm$ 1.0  \\
 NGC 4639  & 32.00  $\pm$ 0.23 & 25.1 $\pm$ 2.5  \\
    &    &    \\
 Mean  & 31.25  $\pm$ 0.20 & 17.8 $\pm$ 1.8    \\
\hline
\end{tabular}

$^1$ N 4536 corrected for ``long" zero point by +0.05 mag
\end{center}
\end{table}

\medskip

Recently we have analyzed a sample  of 37 newly-discovered Cepheids in
the  galaxy NGC~1365 in  the Fornax  cluster  (Silbermann {\it et al.}
1998; Madore {\it et al.} 1998).  These  preliminary results have been
previously reported  in Freedman {\it  et al.} (1997).  Two additional
galaxies  in the Fornax   cluster have  now  been  observed, and   are
currently being analyzed:  NGC 1326A and NGC  1425. The Fornax cluster
is   a particularly important cluster because   it is very compact and
contains galaxies with a range of morphological types.  In contrast to
the Virgo cluster, the small angular size of  the Fornax cluster makes
the determination of its distance much more straightforward.  Thus, it
can provide an important calibration of several secondary methods.  Of
particular interest is the  fact that the  Fornax cluster contains two
well-observed  recent type Ia  supernovae,  which allows for  a direct
comparison between the type Ia  distance scale and other  well-studied
secondary indicators with  small  measured  dispersions, such as   the
Tully-Fisher relation, surface  brightness fluctuations, and planetary
nebula luminosity function.

Correcting for a derived total  line-of-sight reddening of $E(V-I) = $
0.10~mag (derived from the NGC~1365 Cepheids  themselves) gives a true
distance modulus of  $\mu_0 =$ 31.3  $\pm$0.1~mag  (Silbermann {\it et
al.}  1998).  This  corresponds  to  a  distance  to  NGC~1365 of 18.2
$\pm$1.0~Mpc.   This distance  agrees well  with  the distances to the
Fornax cluster determined previously by Jacoby {\it et al.} (1997) and Tonry {\it
et al.} (1997).

A  determination of the Hubble constant  at the distance of the Fornax
cluster requires a knowledge of the Local Group infall velocity to the
Virgo cluster.  Fortunately, however, the derived infall correction to
Fornax  is quite insensitive to the  assumed infall velocity to Virgo:
for an infall velocity of +200 $\pm$100~km/sec the flow correction for
Fornax is only --40  $\pm$20~km/sec, yielding a cosmological expansion
rate of Fornax (determined from the barycentre of  the Local Group) of
1,330~km/sec (Madore {\it et al.}  1998).  Using our Cepheid  distance
of 18.2~Mpc  for Fornax gives $H_0  =$ 73 $\pm$3 (statistical) ~~$\pm$
18 (systematic) km/sec/Mpc.

To conclude this section on local  H$_0$ determinations, an average of
the   six independent determinations  based    on nearby galaxies  and
groups,  including  the Virgo and  Fornax  clusters, gives $H_0  =$ 75
$\pm$15~km/sec (Madore  {\it et al.} 1998).  It   should be noted that
the determinations of $H_0$ in this section make no explicit allowance
for  the  possibility that  the inflow-corrected  velocities of nearby
clusters could be perturbed significantly by other mass concentrations
or large-scale  flows  beyond   the  Virgo cluster.  However,    it is
interesting to note  that these local  estimates agree  very well with
the determinations of     H$_0$ at larger distances,  where   peculiar
velocities are a fractionally-smaller uncertainty (\S3.1,3.2).

\subsection{ Tests of the Cepheid Distance Scale  }

A  number  of  tests  on  the Cepheid   distance  scale are  currently
feasible.  Space here precludes a detailed discussion of many of these
efforts, but   in brief, comparisons of   Cepheid distances with other
independently-calibrated distance indicators  ({\it e.g.,} RR  Lyraes,
and tip  of the red giant branch  (TRGB)) ({\it e.g.,} see the reviews
by Freedman \& Madore 1993; Freedman \& Madore 1996 ; Madore, Freedman
\& Sakai 1997) distances  agree with the  Cepheid distances at a level
of  $\pm$0.1  mag $rms$  (Lee {\it  et al.} 1993;  Sakai  {\it et al.}
1996).   The  agreement between the   absolute calibrations  of the RR
Lyraes (upon which  the TRGB distances  are based) and the Cepheids is
still a matter  of some debate,  with an $rms$ uncertainty of $\pm$0.1
mag.  This uncertainty lies at the heart of the current uncertainty in
the distance modulus to the Large Magellanic Cloud, $\mu$ = 18.5 $\pm$
0.1  mag  ({\it  e.g.,}, see the  review   by  Westerlund 1997),  which
currently   provides  the   fiducial  period-luminosity relations  for
extragalactic distances.

Distances to galaxies based on Cepheids can be  compared on a relative
basis with a  number of other indicators  including the tip of the red
giant branch, planetary nebula luminosity function, surface brightness
fluctuations,  and types I and II  supernovae.  The agreement in these
cases is also quantitatively   very encouraging.  For a  comparison of
recent   Cepheid distances obtained as part   of the H$_0$ Key Project
with   those  based on these other   methods,  see Freedman, Madore \&
Kennicutt (1997).

\subsection{ Is there a Significant Metallicity Dependence? }

A  potentially  important  systematic effect on   the Cepheid distance
scale is the metal abundance.  To date, there has been no consensus on
how significant such an effect might be.  Unfortunately, theory cannot
currently provide  a definitive answer to  the issue  of how abundance
affects  the observed  luminosities   of Cepheids.  Recent  models  by
Chiosi, Wood \& Capitanio (1993) suggest that at the V and I wavebands
observed with HST, the  effect of  abundance amounts to  approximately
--0.1 mag/dex.  Earlier models by  Stothers (1988) and Iben and Tuggle
(1975) were based on B and  V photometry, and  predicted a much larger
effect (see  Freedman \& Madore 1990  for a summary of these results).
According to these models,  both the sense   and the magnitude of  the
effects of   metallicity are dependent  on  wavelength: in   the blue,
higher metallicity Cepheids appear fainter due to line blanketing. The
magnitude effect is smaller in the red, and  the redistribution of the
line-blanketed radiation makes  the  Cepheids  appear brighter.   More
recent work incorporating new opacities, by  Chiosi, Wood \& Capitanio
(in  preparation),  predicts a smaller effect,   0.06 mag/dex, but the
sign is now in the opposite sense.  Hence, the theoretical predictions
are not yet firm and empirical studies are critical for placing limits
on the magnitude of any abundance effects.

At present, the observational  situation also remains unresolved.  The
first observational test  for an abundance  dependence  of the Cepheid
period-luminosity relation   was  undertaken by  Freedman  and  Madore
(1990). These authors observed samples of  Cepheids at three positions
in M31 with respect   to the radial  gradient  in metallicity.    They
concluded that, after correcting for reddening, the difference in true
modulus that  could be attributed   to metallicity was less than  10\%
over a range in abundance of a factor of $\sim$3.
These data were subsequently reanalyzed by Gould
(1994) who concluded,   to  the contrary,   that over a   range in
metallicity of   1  dex,  a  0.56  to  0.88  mag  difference  would be
measurable.

The  results of  Gould (1994), however,  are  inconsistent with  other
limits from  comparisons     of Cepheid  distances     with completely
independent distance    methods  such as  the    TRGB method discussed
above.  For a wide range of  both TRGB and  Cepheid metallicities, the
relative  distances  between  the two   methods   agree to within  the
1-$\sigma$ uncertainties of each of the  methods. As stated above, the
$rms$ differences amount to less than 0.1 mag.

More recently, Sasselov {\it et al.} (1997) and  Beaulieu {\it et al.}
(1997) have analyzed data taken as part of the  EROS search for MACHOS
in  the LMC and SMC.  They  find a dependence of --0.44$^{+0.1}_{-0.2}$
mag/dex.  This  value is  similar  to the value obtained  by  Kochanek
(1997) based on  an analysis of Cepheid data  for a number of galaxies
obtained from a variety of sources, and  with a variety of bandpasses.
An analysis of Galactic Cepheids by Sekiguchi \& Fukugita (1997) finds
a   much  steeper  dependence,  very strongly  in   conflict  with the
constraints provided by other distance indicators.

As part of the Key Project, we have undertaken a second empirical test
in two  fields in  the  face-on spiral galaxy M101.    We find a small
dependence  on metallicity, again  with  a large uncertainty:  $\Delta
\mu_0$ / $\Delta$ [O/H] = --0.24  $\pm$ 0.16 mag/dex (Kennicutt {\it et
al.}  1998).  Again, comparison  of the  Key Project Cepheid distances
with  other  distance indicators provides a  strong  constraint on the
size of  a metallicity  effect.  The metallicity   of the galaxies for
which Cepheid   searches have been  undertaken span  a  range in [O/H]
abundance of almost an order of magnitude, with a median value of --0.3
dex. The  Large    Magellanic  Cloud, which   currently  provides  the
calibrating  period-luminosity relation  for extragalactic  distances,
has  a  very similar abundance  of  [O/H] = --0.4   dex.  These results
suggest that in individual cases  the metallicity effect may amount to
10\%, but the overall effect  on the calibration of secondary distance
indicators will be less than a few percent.

Further progress on constraining  the size of a metallicity dependence
will come   from  NICMOS  observations of  Cepheids   with a  range of
metallicities, currently scheduled  for  this upcoming HST  cycle.  At
long  wavelengths, the   reduced sensitivity  to  both reddening   and
metallicity will improve the accuracy  in the resulting distances by a
factor of $>$2.


\subsection{ Recent Results from Hipparcos }

Feast and Catchpole (1997)  have recently published the first  results
on parallaxes  to  Galactic Cepheids based  on  measurements  from the
Hipparcos satellite.  Based on data for the 26 highest signal-to-noise
Cepheid  parallaxes,  they calibrate the    zero point of  the  V-band
Galactic  Cepheid period-luminosity relation,  adopting the slope from
prior     work      on      LMC     Cepheids.    Correcting        for
$E(B-V)_{LMC}~=~0.074$~mag,     adding  a  theoretical     metallicity
correction  of +0.042~mag, and  adopting $<V>_o-~log(P)$ from Caldwell
\&    Laney     (1991),    they      derive     a    distance  modulus
$(m-M_V)_o^{LMC}~=~18.70\pm0.10$~mag, based on the V-band PL relation.

Madore  \& Freedman   (1997)    have  also  calibrated  the    Cepheid
period-luminosity  relation based   on  the Hipparcos  parallaxes  for
Galactic  Cepheids published by Feast \&  Catchpole (1997), but at six
wavelengths (BVIJHK).  Unfortunately,  the current parallax errors for
the   fundamental    pulsators are    very   large (they     range  in
signal-to-noise =  $\pi /\sigma_{\pi}$ from 0.3 to   5.3, at best) and
they  preclude   an    unambiguous  interpretation  of    the observed
differences.   These differences may  arise from a combination of true
distance modulus,   reddening and/or  metallicity  effects.  Currently
extragalactic distances are calibrated  relative to those of the Large
Magellanic Cloud (LMC).  These results  suggest a range of LMC  moduli
between  18.44 $\pm$ 0.35  and 18.57  $\pm$ 0.11 mag   (49 to 52 kpc).
Comparing these calibrations with previously published multiwavelength
PL relations   from Madore  \&  Freedman  (1991), there is   very good
agreement at   a level of  0.07  $\pm$ 0.14 mag,   or 4 $\pm$   7\% in
distance.  Madore and  Freedman adopted 18.50 $\pm$  0.10 mag for  the
distance to the LMC.

Recently,  there have been a  number of other independent measurements
of the distance to the LMC.  A  new, independent measurement of the RR
Lyrae distance  to the LMC yields  18.48$\pm$ 0.19 mag (Alcock {\it et
al.} (1997). Based  on an analysis of  the expanding ring of supernova
1987A, Gould \& Uza  (1997) derive $\mu_{LMC} <$  ~18.37 $\pm0.04$~mag
for the  LMC  true  distance modulus if  the   ring is assumed  to  be
circular; they   note that if  the  ring is slightly  elliptical ($b/a
\sim$  0.95) this upper limit increases  to $<~18.44~\pm~0.05$~mag.  A
value of 18.56 $\pm$ 0.05~mag has been derived by Panagia {\it et al.}
(1996) from the same data. For the present time, the H$_0$ Key Project
has adopted a  true distance modulus of 18.50  $\pm$ 0.10 mag for  the
LMC.  This  value is consistent  with other estimated distances to the
LMC based on a wide variety of methods ({\it e.g.,} Westerlund 1997).




\section{ Summary  }

Fifteen   years ago  at   the 1982  General   Assembly, there   was no
reconciling   the  values   of  H$_0$   obtained  from   the  infrared
Tully-Fisher relation presented by the late  Marc Aaronson (H$_0$ = 85
$\pm$ 5 km/sec/Mpc; Aaronson 1983), the value of 50 $\pm$ 7 km/sec/Mpc
presented by Gustav Tammann and Allan Sandage  based primarily on type
Ia supernovae calibrated by M supergiants via  Cepheids, and the value
of  H$_0$ =  95 $\pm$  10 km/sec/Mpc derived   by  the late Gerard  de
Vaucouleurs,  based on  a number  of   different methods.  At  the IAU
Symposium  183  in 1997, we appear  to  be  seeing some convergence in
values of H$_0$, with  values of 55  $\pm$ 10 from type Ia  supernovae
being reported by Gustav Tammann, and 73 $\pm$ 6 (statistical) $\pm$ 8
(systematic) km/sec/Mpc by  our   group (and  based  on a   number  of
independent secondary  methods).  Perhaps  most importantly, the error
bars  now overlap and all  groups are now  quoting both statistical as
well as systematic errors.

Our systematic error takes into account a number of factors including:
the    present  uncertainty  in    the  zero  point  of    the Cepheid
period-luminosity relation of $\pm$5\% (effectively the uncertainty in
the distance to  the LMC), the uncertainty  due to metallicity in  the
Cepheid period-luminosity   relation   at  a level   of  $\pm$5\%,  an
uncertainty  of $\pm$7\%  that   allows for the  possibility  that the
locally measured  H$_0$  out to $\sim$10,000  km/sec   may not  be the
global  value of H$_0$,  plus an  allowance  for a scale error  in the
photometry that  could  affect all  of  the results  at the   level of
$\pm$3\%.   At  the present time,   the total uncertainties  amount to
about   $\pm$15\%.  This  result is  based   on a variety of  methods,
including a Cepheid calibration of  the Tully-Fisher relation, type Ia
supernovae, a  calibration  of distant clusters  tied   to Fornax, and
direct Cepheid distances out to $\sim$ 20~Mpc.

In the next couple of  years, all of  the observations and analysis of
H$_0$ Key Project galaxy sample will have been undertaken, and a final
calibration of secondary  distance indicators  can  be completed.   In
addition, new, near-infrared  H-band (1.6 $\mu$m) NICMOS  observations
are now being scheduled on HST that will minimize the dominant sources
of   systematic uncertainty   in   the  Cepheid distances   (currently
reddening and metallicity).  New, optical  and infrared photometry are
being obtained for the Cepheid Tully-Fisher calibrators.  New data are
being obtained  for the distant  Tully-Fisher  galaxies.  Ground-based
studies   are dramatically  increasing  the  numbers  of well-observed
supernovae.  There are now quantitative reasons  for optimism that the
extragalactic  distance scale will soon  be  firmly established at the
$\pm$10\% level.

\section{Acknowledgments}

We  are  pleased  to   acknowledge   the   enormous  efforts  of   our
co-investigators  on the H$_0$ Key   Project  team: L.  Ferrarese,  H.
Ford, J.  Graham,  M.  Han, J.    Hoessel, J.  Huchra,  S.  Hughes, G.
Illingworth, R.  Phelps,  A.  Saha, S.   Sakai, N.  Silbermann, and P.
Stetson, and graduate students F.  Bresolin,  P.  Harding, D.  Kelson,
L.  Macri, D.  Rawson,   and  A. Turner.    This   work is based    on
observations with the NASA/ESA Hubble Space Telescope, obtained by the
Space Telescope  Science Institute, which  is  operated by  AURA, Inc.
under NASA contract No.  5-26555.  Support for  this work was provided
by NASA through grant GO-2227-87A from STScI.




\vfill\eject




\end{document}